\documentclass[aps,pre,twocolumn,10pt,showpacs,superscriptaddress,groupedaddress,longbibliography]{revtex4-1} 
\pdfoutput=1

\usepackage{graphicx}        
\usepackage{dcolumn}        
\usepackage{bm}             
\usepackage{amssymb}
\usepackage{amsmath}
\usepackage{epstopdf} 
\usepackage{color}
\definecolor{korr_26Apr}{rgb}{0,0,0} 
\definecolor{red}{rgb}{1,0,0}

\def \d{\mathrm{d}}

\definecolor{red}{rgb}{1,0,0}

\begin{document}

\widetext

\title{Fluid forces or impacts: What governs the entrainment of soil particles in sediment transport mediated by a Newtonian fluid?}
\author{Thomas P\"ahtz$^{1,2}$}
\email{0012136@zju.edu.cn}
\author{Orencio Dur\'an$^3$}
\affiliation{1.~Institute of Port, Coastal and Offshore Engineering, Ocean College, Zhejiang University, 866 Yu Hang Tang Road, 310058 Hangzhou, China \\
2.~State Key Laboratory of Satellite Ocean Environment Dynamics, Second Institute of Oceanography, 36 North Baochu Road, 310012 Hangzhou, China \\
3.~Virginia Institute of Marine Science, College of William and Mary, 1375 Greate Road, Gloucester Point, Virginia 23062, USA}

\begin{abstract}
In steady sediment transport, the deposition of transported particles is balanced by the entrainment of soil bed particles by the action of fluid forces or particle-bed impacts. Here we propose a proxy to determine the role of impact entrainment relative to entrainment by the mean turbulent flow: the ``bed velocity'' $V_b$, which is an effective near-bed-surface value of the average horizontal particle velocity that generalizes the classical slip velocity, used in studies of aeolian saltation transport, to sediment transport in an arbitrary Newtonian fluid. We study $V_b$ for a wide range of the particle-fluid-density ratio $s$, Galileo number $\mathrm{Ga}$, and Shields number $\Theta$ using direct sediment transport simulations with the numerical model of Dur\'an et al. (Phys. Fluids \textbf{24}, 103306, 2012), which couples the discrete element method for the particle motion with a continuum Reynolds-averaged description of hydrodynamics. We find that transport is fully sustained through impact entrainment (i.e., $V_b$ is constant in natural units) when the ``impact number'' $\mathrm{Im}=\mathrm{Ga}\sqrt{s+0.5}\gtrsim20$ or $\Theta\gtrsim5/\mathrm{Im}$. These conditions are obeyed for the vast majority of transport regimes, including steady turbulent bedload, which has long been thought to be sustained solely through fluid entrainment. In fact, we find that transport is fully sustained through fluid entrainment (i.e., $V_b$ scales with the near-bed horizontal fluid velocity) only for sufficiently viscous bedload transport at grain scale (i.e., for $\mathrm{Im}\lesssim20$ and $\Theta\lesssim1/\mathrm{Im}$). Finally, we do not find a strong correlation between, $V_b$, or the classical slip velocity, and the transport-layer-averaged horizontal particle velocity $\overline{v_x}$, which challenges the long-standing consensus that predominant impact entrainment is responsible for a linear scaling of the transport rate with $\Theta$. For turbulent bedload in particular, $\overline{v_x}$ increases with $\Theta$ despite $V_b$ remaining constant, which we propose is linked to the formation of a liquid-like bed on top of the static-bed surface.

\end{abstract}

\maketitle

\section{Introduction}
Sediment transport in a Newtonian fluid, such as water and air, is one of the most important geological processes responsible for the alteration of sea and riverscapes, and dry planetary surfaces \cite{Yalin77,Graf84,vanRijn93,Julien98,Garcia07,Bourkeetal10,Bagnold41,Shao08,Duranetal11,Koketal12,Valanceetal15}. It can occur in a large variety of natural environments: e.g., viscous and turbulent transport of minerals and organics by Earth's water streams \cite{Yalin77,Graf84,vanRijn93,Julien98,Garcia07}, and turbulent transport of dust and sand by Earth's atmospheric winds \cite{Bagnold41,Shao08,Duranetal11,Koketal12,Valanceetal15}. 

Different sediment transport regimes are documented. Very small particles, whose weight can be fully supported by the fluid turbulence, tend to be transported in turbulent suspensions \cite{Garcia07,Koketal12}. Medium and large particles, on the other hand, are transported close to the surface, in trajectories not much influenced by fluid turbulence \cite{Garcia07,Koketal12}. The latter case includes bedload and saltation transport. Bedload transport refers to particles rolling, sliding, and hopping in the vicinity of the sediment bed, which is typical for the transport of sand and gravel by water streams \cite{FreyChurch09,FreyChurch11}. Saltation transport refers to particles moving in ballistic trajectories along the bed, which is typical for the transport of sand by planetary winds \cite{Duranetal11,Koketal12,Valanceetal15}.

It has become a widely accepted hypothesis that the mechanisms sustaining bedload and saltation transport are fundamentally different: bedload transport being sustained through entrainment of soil bed particles directly by fluid forces \cite{Shields36,Ward69,VanRijn84a,BuffingtonMontgomery97,Wilcock98,Buffington99,Dey99,Dey03,Paphitis01,Caoetal06,VollmerKleinhans07,BeheshtiAtaieAshtiani08,DeyPapanicolaou08,Recking09,Diplasetal08,Valyrakisetal10,Valyrakisetal13,FreyChurch11,Leeetal12,Houssaisetal15,AliDey16,Maniatisetal17,Rousaretal16} and saltation transport being sustained through particle-bed impacts ejecting bed particles \citep{Bagnold37,Bagnold73,UngarHaff87,AndersonHaff88,AndersonHaff91,McEwanWillets91,McEwanWillets93}, allowing transport even below the fluid entrainment threshold \cite{Kok10a,Kok10b,Duranetal11,Paehtzetal12,Koketal12,JenkinsValance14,Carneiroetal15,Valanceetal15,Berzietal16,Berzietal17,PaehtzDuran17b}. However, recent studies have questioned this hypothesis by pointing out the role of particle inertia for sustaining bedload transport \cite{Clarketal15,Clarketal17}.

Here we study the relevance of impact entrainment relative to direct entrainment by the mean turbulent flow in a unified manner using direct sediment transport simulations in a Newtonian fluid with the model of Ref.~\cite{Duranetal12}, which belongs to a new generation of sophisticated grain-scale models of sediment transport \cite{Carneiroetal11,Duranetal11,Duranetal12,Carneiroetal13,Jietal13,Duranetal14a,Duranetal14b,KidanemariamUhlmann14a,KidanemariamUhlmann14b,KidanemariamUhlmann17,Schmeeckle14,Vowinckeletal14,ArollaDesjardins15,Paehtzetal15a,Carneiroetal15,Clarketal15,Derksen15,Maurinetal15,Maurinetal16,FinnLi16,Finnetal16,PaehtzDuran17b} and has been shown to reproduce many observations concerning viscous and turbulent sediment transport in air and water \cite{Duranetal11,Duranetal12,Duranetal14a,PaehtzDuran17b} (e.g., see Fig.~6 of Ref.~\cite{Duranetal14a} and Figs.~2, 6, and 7 of Ref.~\cite{PaehtzDuran17b}), and bedform formation \cite{Duranetal14b}. First, we present direct evidence from visual inspection of these simulations showing that impact entrainment events play a crucial role during both steady bedload and saltation transport. Then we quantitatively analyze the relative role of impact entrainment from our simulation data using a proxy that is similar, but not identical, to the classical slip velocity (i.e., the average horizontal particle velocity at the bed surface \cite{Andreotti04,Ogeretal05,ClaudinAndreotti06,Beladjineetal07,Ogeretal08,Kok10b,Duranetal11,Paehtzetal12,JenkinsValance14,Berzietal16,Berzietal17}). This analysis reveals a crucial influence of a dimensionless number, henceforth called ``impact number'': $\mathrm{Im}=\mathrm{Ga}\sqrt{s+0.5}$, where $s=\rho_p/\rho_f$ is the particle-fluid-density ratio and $\mathrm{Ga}=\sqrt{(s-1)gd^3}/\nu$ the Galileo number (the square root of the Archimedes number), with $g$ the gravitational constant, $d$ the mean particle diameter, and $\nu$ the kinematic viscosity. Finally, we shed light on possible links between impact entrainment and average transport characteristics, such as the scaling of the sediment transport rate $Q$ with the dimensionless fluid shear stress (the ``Shields number'' $\Theta=\tau/[(\rho_p-\rho_f)gd]$). In fact, it is a widespread belief that impact entrainment inevitably causes a (nearly) linear scaling $Q\propto\Theta-\Theta^r_t$, where $\Theta^r_t$ is the extrapolated value of $\Theta$ at which $Q$ vanishes \cite{UngarHaff87,Andreotti04,Kok10a,Kok10b,Paehtzetal12,Laemmeletal12,JenkinsValance14,Berzietal16,Berzietal17,PaehtzDuran17b}. Here we challenge this belief.

The reminder of the paper is organized as follows. Section~\ref{Simulations} briefly summarizes important details of the numerical model and explains how we calculate average quantities from the simulation data. Section~\ref{VisualizationsImpact} presents and discusses the evidence for impact entrainment in both bedload and saltation transport obtained from visualizations of the numerical simulations. Section~\ref{Proxy} represents the core of the paper. It introduces the proxy we use to quantify the relative role of impact entrainment, explains why other proxies, such as the slip velocity, are inappropriate, and analyzes our proxy over the entire range of simulated conditions. Section~\ref{Implications} discusses possible links between impact entrainment and average transport characteristics. Finally, we discuss our results and draw conclusions in Sec.~\ref{Conclusions}.

\section{Numerical Simulations} \label{Simulations}
The numerical model of sediment transport in a Newtonian fluid of Ref.~\cite{Duranetal12} couples a discrete element method for the particle motion ($\approx15000$ spheres, including $>10$ layers of sediment bed particles) with a continuum Reynolds-averaged description of hydrodynamics. The Reynolds-Averaged Navier-Stokes equations are combined with an improved mixing length approximation, which can be used to calculate the mean turbulent fluid velocity at high particle concentrations. In contrast to the original model, which considers only gravity, buoyancy, and fluid drag forces acting on particles, we here also consider the added-mass force \cite{Duranetal11}. However, cohesive and higher-order fluid forces, such as the hindrance and lift force remain neglected. We also corrected two slight inaccuracies in the original model (with a mostly negligible effect on the simulation outcome): We here take into account that the fluid shear stress is proportional to the fluid volume fraction and neglect the buoyancy contribution from the divergence of the fluid shear stress because the divergence of the Reynolds stress, previously considered, actually does not contribute to the buoyancy force.

We would also like to emphasize that all results presented in this study for the bedload transport regime usually do not significantly depend on contact parameters, such as the restitution coefficient $e$ and contact friction coefficient $\mu^c$. For instance, bedload transport simulations with $e=0.9$ and $e=0.01$ are nearly exactly the same on average, which is consistent with previous reports \cite{Maurinetal15}. This finding implies that any dissipative interaction force that is proportional to the relative velocity of two approaching particles and acts at and/or close to particle contact, such as the lubrication force \cite{SimeonovCalantoni12}, does not significantly influence average bedload transport characteristics as the effect of such forces can be incorporated in $e$ and $\mu_c$ \cite{Gondretetal02,YangHunt06,Schmeeckle14,Maurinetal15}.

We carry out simulations for $s$ and $\mathrm{Ga}$ within the range $s\in[1.1,10^7]$ and $\mathrm{Ga}\in[0.1,100]$. For each pair of $s$ and $\mathrm{Ga}$, we vary $\Theta$ in regular intervals above the entrainment cessation threshold $\Theta^e_t$, which is usually larger than the rebound cessation threshold $\Theta^r_t$ associated with vanishing sediment transport \cite{PaehtzDuran17b}. We use the simulation data to compute local and transport layer averages of particle and fluid properties, such as the particle stress tensor, which is explained in the following.

\subsection{Local, mass-weighted ensemble average}
We compute the local, mass-weighted ensemble average $\langle A\rangle$ of a particle quantity $A$ through \cite{Paehtzetal15a}
\begin{eqnarray}
 \langle A\rangle&=&\frac{1}{\rho}\overline{\sum_nm^nA^n\delta(\mathbf{x}-\mathbf{x}^n)}^E, \\
 \rho&=&\overline{\sum_nm^n\delta(\mathbf{x}-\mathbf{x}^n)}^E, 
\end{eqnarray}
where $\rho$ is the local particle mass density, $m$ is the particle mass, $\delta$ is the $\delta$ distribution, and $\mathbf{x}=(x,y,z)$ (Cartesian coordinate system) is the location, with $x$ in the flow direction parallel to the bed, $z$ in the direction normal to the bed oriented upwards, and $y$ in the lateral direction. Furthermore, the sum iterates over all particles ($n\in(1,N)$, with $N$ the total number of particles), and $\overline{\cdot}^E$ denotes the ensemble average.

\subsection{Particle stress tensor}
Using the definition of the local mass-weighted ensemble average, we compute the particle stress tensor $P_{ij}$ from the simulation data through \cite{Paehtzetal15a}
\begin{eqnarray}
 P_{ij}&=&\rho\langle v^\prime_iv^\prime_j\rangle+\frac{1}{2}\overline{\sum_{mn}F_j^{mn}(x^m_i-x^n_i)K(\mathbf{x},\mathbf{x}^m,\mathbf{x}^n)}^E, \nonumber \\
 && \\
 \mathbf{v^\prime}&=&\mathbf{v}-\langle\mathbf{v}\rangle, \nonumber \\
 K&=&\int\limits_0^1\delta\{\mathbf{x}-[(\mathbf{x}^m-\mathbf{x}^n)s^\prime+\mathbf{x}^n]\}\d s^\prime, \nonumber
\end{eqnarray}
where $s^\prime$ is a dummy variable, $\mathbf{v}$ is the particle velocity, and $\mathbf{F}^{mn}$ is the contact force applied by particle $n$ on particle $m$ ($\mathbf{F}^{mm}=0$).

\subsection{Transport layer average}
We compute the transport layer average $\overline{A}$ of a quantity $A$ through
\begin{eqnarray}
 \overline{A}=\int_{z_r}^\infty\rho\langle A\rangle\d z/\int_{z_r}^\infty\rho\d z.
\end{eqnarray}
It describes a mass-weighted average of $A$ over all particles within the transport layer ($z>z_r$), where the transport layer base height $z_r$ is defined through
\begin{eqnarray}
 \max(P_{zz}\dot\gamma)=[P_{zz}\dot\gamma](z_r), \label{ShearWork}
\end{eqnarray}
with $\dot\gamma=\d\langle v_x\rangle/\d z$ the particle shear rate. This definition is motivated by the fact that the term $P_{zz}\dot\gamma$ is the production rate of the cross-correlation fluctuation energy density $\rho\langle(v^\prime_xv^\prime_z)\rangle$ \cite{Paehtzetal15a}. Because particle-bed rebounds are the main reason for the production of $\rho\langle(v^\prime_xv^\prime_z)\rangle$, since they effectively convert the horizontal momentum of descending particles into the vertical momentum of ascending particles, $z_r$ is a measure for the effective location of energetic particle-bed rebounds.

\begin{figure}[htb]
 \begin{center}
  \includegraphics[width=0.8\columnwidth]{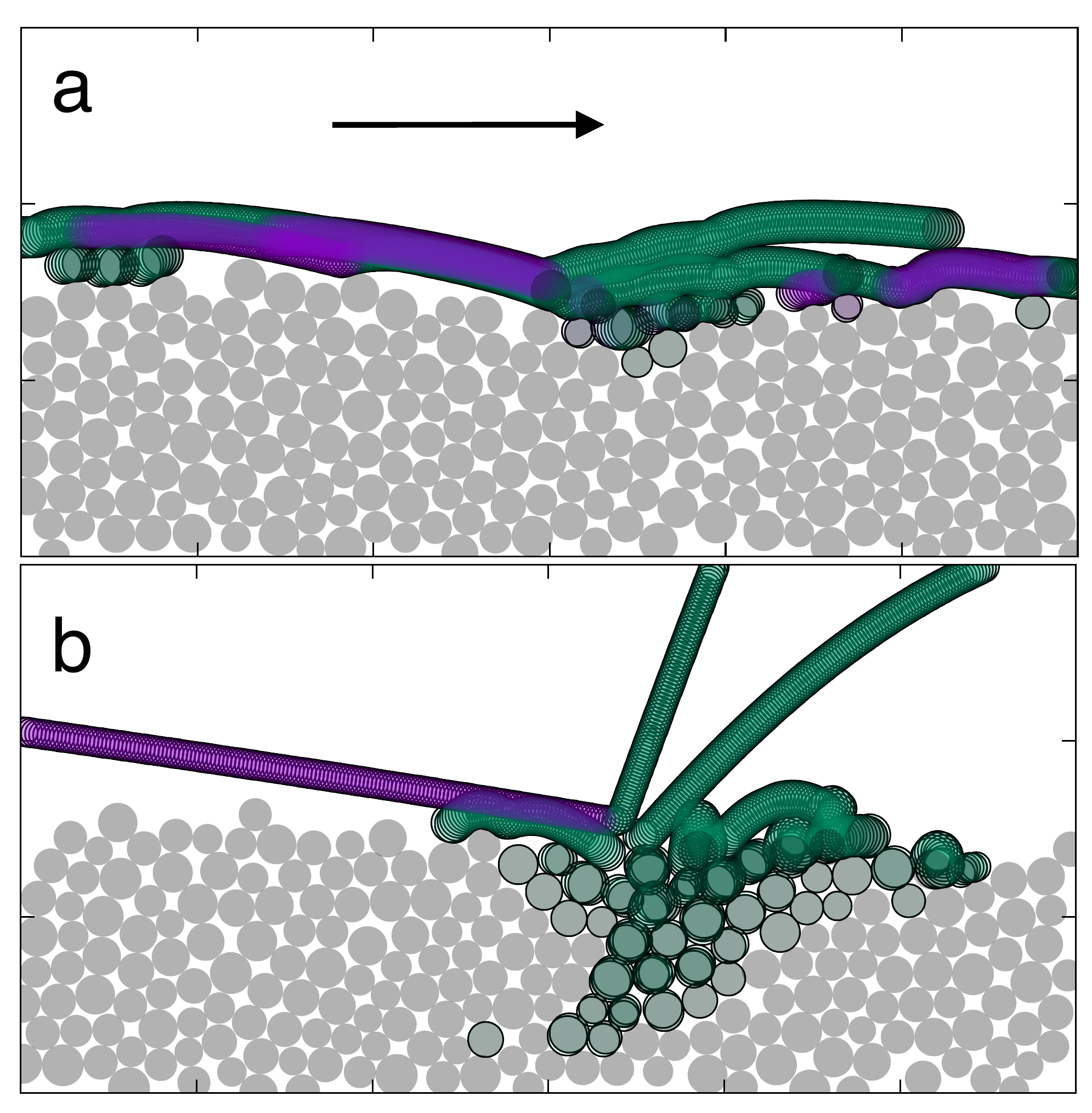}
 \end{center}
 \caption{Visualizations of impact entrainment events for (a) turbulent bedload and (b) saltation transport from direct sediment transport simulations near threshold conditions ($\Theta\approxeq\Theta^e_t$). The purple and green colors indicate the trajectory of particles before and after impact, respectively. Particles move from the left to the right.}
 \label{Visualizations}
\end{figure}
\begin{figure*}[htb]
 \begin{center}
  \includegraphics[width=2.0\columnwidth]{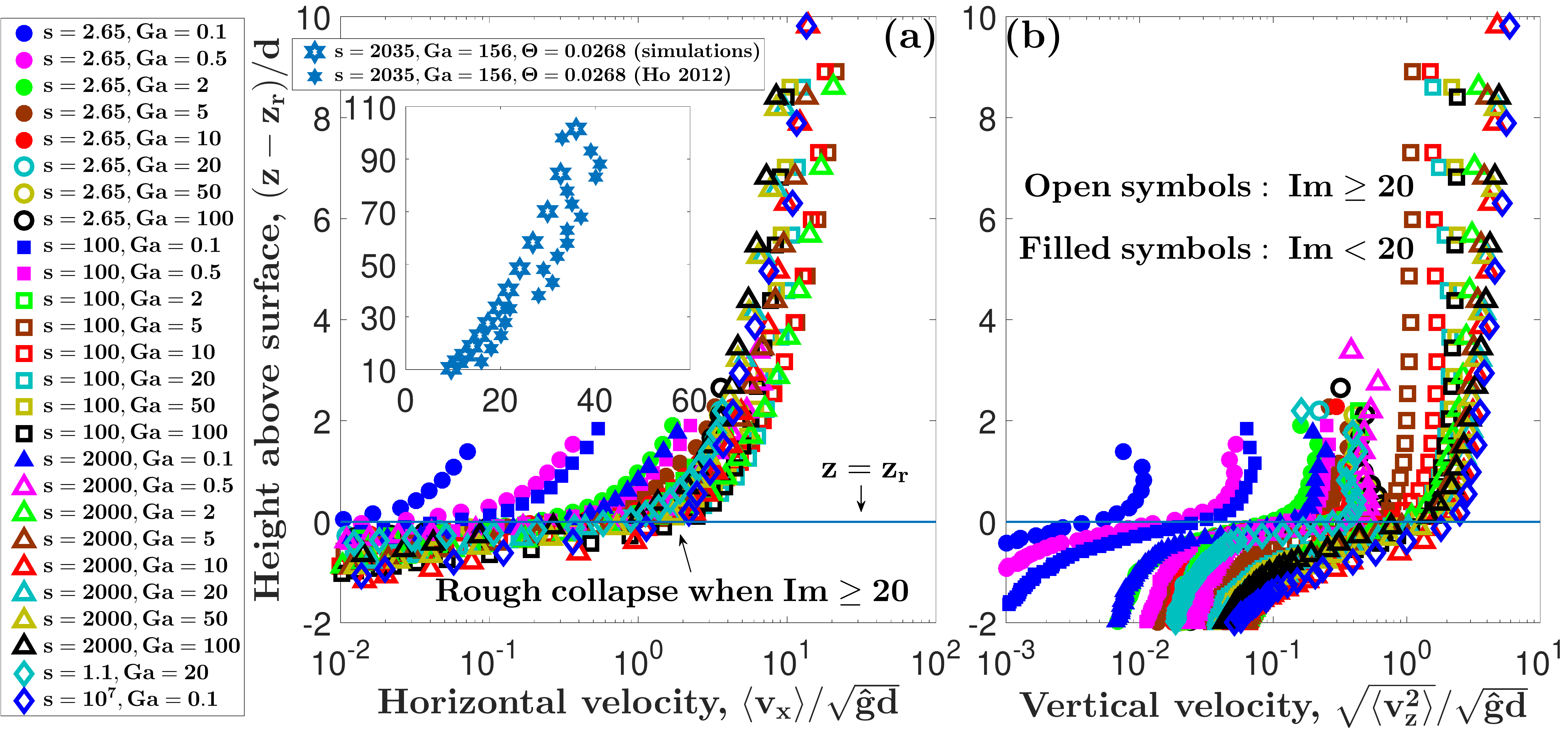}
 \end{center}
 \caption{Vertical profiles of (a) $\langle v_x\rangle/\sqrt{\hat gd}$ and (b) $\sqrt{\langle v_z^2\rangle/(\hat gd)}$ for various $s$ and $\mathrm{Ga}$ near threshold conditions ($\Theta\approxeq\Theta^e_t$). Inset of (a): Comparison of simulated vertical profile of $\langle v_x\rangle(z_r)/\sqrt{\hat gd}$ with a profile measured in a wind tunnel by Ref.~\cite{Ho12} for coarse sand ($d=630\;\mu$m).}
 \label{Velprofiles}
\end{figure*}

\section{Visualizations of impact entrainment events} \label{VisualizationsImpact}
Figure~\ref{Visualizations} shows snapshots of transport simulations near $\Theta^e_t$ for turbulent bedload transport [Fig.~\ref{Visualizations}(a)] and saltation transport [Fig.~\ref{Visualizations}(b)] shortly before and after an impact entrainment event (see also Movies~S1 and S2 in the Supplementary Material \cite{Suppl_Mat}). In both cases, a transported particle collides with one or more bed particles, which subsequently become mobilized with a short time delay. In bedload transport, the transported particle impacts the bed from a very small height and entrains a bed particle by dragging it out of its trap, while in saltation transport, the transported particle impacts the bed from a very large height and entrains a bed particle by ejecting it. Close to the threshold, impact entrainment events are rare in both cases as impacts occur much less often, but are much more effective, in saltation than in turbulent bedload transport (Movies~S1 and S2 \cite{Suppl_Mat}). As a consequence, bed particles remain in repose most of the time as fluid forces are too weak to entrain them directly. Sufficiently far from the threshold, impact entrainment events occur much more often in saltation transport (Movie~S3 \cite{Suppl_Mat}), whereas it is impossible to determine single entrainment events in turbulent bedload transport because several layers of the bed are in continuous motion (Movie~S4 \cite{Suppl_Mat}). Note that movie captions are provided in the Appendix.

\section{Proxy for relevance of impact entrainment relative to fluid entrainment} \label{Proxy}
The finding from the previous section that impact entrainment events play crucial roles during both saltation and turbulent bedload transport highlights the need for a proxy conveying information about the relevance of impact entrainment relative to direct entrainment by the mean turbulent flow. Here we discuss two potential proxies, which we obtain from the vertical profile of the average horizontal particle velocity $\langle v_x\rangle(z)$. Section~\ref{SlipVel} discusses the standard proxy, the slip velocity $\langle v_x\rangle(z_r)$, and why it is inappropriate for our purposes. Therefore, Sec.~\ref{BedVel} proposes and quantitatively analyzes an improved proxy, the ``bed velocity''.

\subsection{Slip velocity} \label{SlipVel}
\begin{figure*}[htb]
 \begin{center}
  \includegraphics[width=2.0\columnwidth]{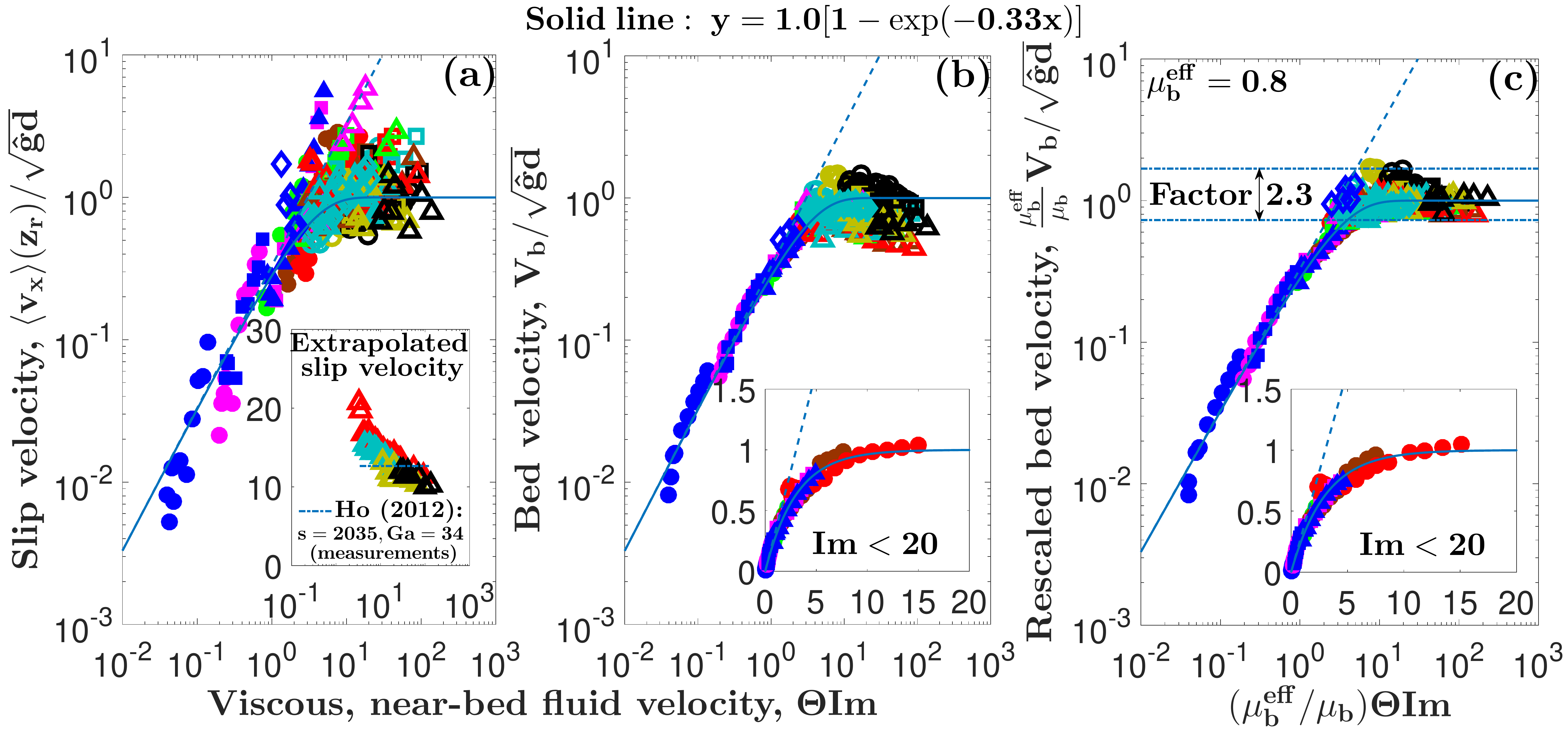}
 \end{center}
 \caption{(a) Dimensionless slip velocity $\langle v_x\rangle(z_r)/\sqrt{\hat gd}$ and (b) dimensionless bed velocity $V_b/\sqrt{\hat gd}$ vs dimensionless, viscous, horizontal near-bed fluid velocity $\Theta\mathrm{Im}$ for various $s$, $\mathrm{Ga}$, and $\Theta$ in log-log scale. (c) Same as (b), but both axes are further rescaled by $\mu_b/0.8$. Inset of (a): Estimate of $\langle v_x\rangle(z_r)/\sqrt{\hat gd}$ obtained from linear extrapolation of $\langle v_x\rangle(z)$ from $z\in(15,110)d$ to $z_r$ vs $\Theta\mathrm{Im}$ for conditions with $s=2000$, $\mathrm{Ga}\geq10$, and various $\Theta$, and comparison to measurements by Ref.~\cite{Ho12}. Insets of (b and c): Same as (b) and (c), but in linear-linear scale for cases with $\mathrm{Im}<20$. For symbol legend, see Fig.~\ref{Velprofiles}.}
 \label{Slipvelocity}
\end{figure*}
Figure~\ref{Velprofiles} shows the vertical profiles of (a) $\langle v_x\rangle/\sqrt{\hat gd}$ and (b) $\sqrt{\langle v_z^2\rangle/(\hat gd)}$ relative to $z_r$ for the entire simulated range of $s$ and $\mathrm{Ga}$, and a value of $\Theta$ that is near the associated entrainment threshold $\Theta^e_t(s,\mathrm{Ga})$, where $\hat g=(s+0.5)g/(s-1)$ is the value of $g$ reduced by the buoyancy and added-mass force. It can be seen that there is a very rough tendency of $\langle v_x\rangle/\sqrt{\hat gd}$, but not of $\sqrt{\langle v_z^2\rangle/(\hat gd)}$, to collapse near $z_r$ when the impact number $\mathrm{Im}\gtrsim20$ (open symbols). In the aeolian research community, a roughly constant value of $\langle v_x\rangle(z_r)/\sqrt{\hat gd}$ is thought to be evidence that saltation transport is a fully impact-sustained transport regime as it is associated with a constant average outcome of particle-bed impacts \cite{Andreotti04,Ogeretal05,ClaudinAndreotti06,Beladjineetal07,Ogeretal08,Kok10b,Duranetal11,Paehtzetal12,Koketal12,JenkinsValance14,Berzietal16,Berzietal17}. In fact, when impact entrainment dominates fluid entrainment, every particle trapped at the bed must be replaced by precisely one particle entrained through impacts on average. However, this line of reasoning is not entirely accurate because it indirectly assumes that all particle-bed impacts occur at the same vertical location $z_r$. However, particle-bed impacts actually occur at varying vertical locations and their range of influence often involves several layers of the sediment bed (e.g., see Movie~S3 \cite{Suppl_Mat}), which makes this assumption problematic because $\langle v_x\rangle(z)$ increases exponentially with $z$ near $z_r$ [Fig.~\ref{Velprofiles}(a)], meaning small changes of $z$ have large effect. Indeed, Fig.~\ref{Slipvelocity}(a) shows that the slip velocity exhibits significant fluctuations with $\Theta\mathrm{Im}$, and thus with $\Theta$, even for typical saltation transport conditions in Earth's atmosphere (e.g., $s=2000$, $\mathrm{Ga}\geq10$). Note that $\Theta\mathrm{Im}=u_\ast^2d/(\nu\sqrt{\hat gd})$, where $u_\ast=\sqrt{\tau/\rho_f}$, is the fluid shear velocity, is the viscous, horizontal near-bed fluid velocity in natural units ($\sqrt{\hat gd}$).

In the aeolian research community, it is the current consensus point of view that experiments (e.g., \cite{Zouetal01,Dongetal04b,Yangetal07,Kangetal08,RasmussenSorensen08,Creysselsetal09,Hoetal11,Ho12}) show an approximately constant slip velocity for saltation transport \cite{Koketal12}, which would contradict our numerical finding if true. However, what the experiments truly show is instead that the {\em{extrapolation}} of $\langle v_x\rangle(z)$ from vertical locations $z\gtrsim z_r+5\;$mm to $z_r$ is approximately constant. In fact, reliable measurements of $\langle v_x\rangle(z)/\sqrt{\hat gd}$ do not exist for vertical locations $z\lesssim z_r+5\;$mm because large particle concentrations near $z_r$ strongly disturb the measurement apparatuses \cite{Ho12}, which is supported by the fact that the few actual measurements (i.e., nonextrapolations) of $\langle v_x\rangle(z)/\sqrt{\hat gd}$ reported for that region \cite{Zouetal01,Dongetal04b,RasmussenSorensen08,Creysselsetal09} vary by more than an order of magnitude between about $0.5$ \cite{Dongetal04b} and $30$ \cite{RasmussenSorensen08}. As shown in the inset of Fig.~\ref{Velprofiles}a, the data for $z\gtrsim z_r+5\;$mm, where our simulations are consistent with measurements, suggest a linear trend of $\langle v_x\rangle(z)$ with $z$ even though the actual trend for $z\lesssim z_r+5\;$mm is much closer to an exponential behavior. It explains why the extrapolation of $\langle v_x\rangle(z)$ to $z_r$ yields values very different from the actual slip velocity. Indeed, when we estimate the slip velocity from our transport simulations via linear extrapolation, we obtain values that are consistent with the likewise extrapolated measurements [inset of Fig.~\ref{Slipvelocity}(a)].

Though the extrapolation of $\langle v_x\rangle(z)/\sqrt{\hat gd}$ to $z_r$ might serve as a proxy for the relative relevance of impact entrainment for saltation transport in Earth's atmosphere, which is characterized by a large transport layer, it is obviously meaningless for transport regimes with a small transport layer, such as bedload transport, and thus does not allow a unified treatment of all transport regimes. In what follows, we therefore propose an improved proxy. 

\subsection{Bed velocity} \label{BedVel}
As explained above, the main issue with the slip velocity proxy is the fact that particle-bed impacts occur at varying vertical locations, rather than at a fixed vertical location $z_r$, and their range of influence often involves several layers of the sediment bed. To mend this issue, we here motivate the definition of an improved proxy, the bed velocity $V_b$, as an effective bed surface value of $\langle v_x\rangle$ that involves several layers around $z_r$.

First, we use that $\langle v_x\rangle$ exponentially decays within the sediment bed with a characteristic decay height proportional to $d$ [Fig.~\ref{Velprofiles}(a)]. An effective value of $\langle v_x\rangle$ must thus be proportional to an effective value $V^\prime_b$ of the horizontal velocity gradient $\dot\gamma$ near the bed surface:
\begin{equation}
 V_b\propto V^\prime_bd. \label{slipvsgrad}
\end{equation}
Second, we calculate $V^\prime_b$ as the ratio between $[-P_{zx}\dot\gamma](z_r)$, which is a suitable definition of an effective bed surface value of $-P_{zx}\dot\gamma$ [cf. Eq.~(\ref{ShearWork})], and a suitable definition of the bed-surface-averaged particle shear stress $-P_{zx}$, namely,
\begin{equation}
 V^\prime_b = \frac{[-P_{zx}\dot\gamma](z_r)}{\frac{1}{\rho_b}\int_{-\infty}^\infty\frac{\d\rho}{\d z}P_{zx}\d z}, \label{Vbed1}
\end{equation}
where the weight $-\rho_b^{-1}\d\rho/\d z$, with $\rho_b\approx0.58\rho_p$ the value of the particle concentration $\rho$ deep within the bed, is maximal near the bed surface as it vanishes sufficiently within and above the bed. After partial integration, using $-P_{zx}(-\infty)=\tau$ and the horizontal momentum balance $\d P_{zx}/\d z=\rho\langle a_x\rangle$ \cite{Paehtzetal15a}, with $\mathbf{a}$ the particle acceleration due to the action of non-contact forces, Eq.~(\ref{Vbed1}) becomes
\begin{eqnarray}
 V^\prime_b=\frac{[-P_{zx}\dot\gamma](z_r)}{\tau-\frac{1}{\rho_b}\int_{-\infty}^\infty\rho^2\langle a_x\rangle\d z}. \label{Vbed}
\end{eqnarray}
Finally, we obtain the proportionality factor in Eq.~(\ref{slipvsgrad}) by imposing that the bed velocity very roughly equals the classical slip velocity, $V_b\approx\langle v_x\rangle(z_r)$, for the simulated turbulent saltation transport cases ($s=2000$, $\mathrm{Ga}\geq10$). This constraint yields
\begin{eqnarray}
 V_b=0.33d\times\frac{[-P_{zx}\dot\gamma](z_r)}{\tau-\frac{1}{\rho_b}\int_{-\infty}^\infty\rho^2\langle a_x\rangle\d z}. \label{Vbedfinal}
\end{eqnarray}

Figure~\ref{Slipvelocity}(b) shows that $V_b/\sqrt{\hat gd}$ is a much better proxy than $\langle v_x\rangle(z_r)/\sqrt{\hat gd}$ [Fig.~\ref{Slipvelocity}(a)] as it reproduces the approximately constant behavior expected for saltation transport in Earth's atmosphere. In detail, we observe two extreme regimes. When sediment transport is fully sustained through direct entrainment by the mean turbulent flow, the bed velocity scales with the average near-bed fluid velocity [dashed line in Fig.~\ref{Slipvelocity}(b)]. In contrast, when sediment transport is fully sustained through impact entrainment, the dimensionless bed velocity does not change much with $\Theta$, $s$, and $\mathrm{Ga}$: $V_b/\sqrt{\hat gd}\approx 1.0$. A part of the variation of $V_b/\sqrt{\hat gd}$ in this regime can be attributed to small changes of the bed friction coefficient $\mu_b=\mu(z_r)$, where $\mu=-P_{zx}/P_{zz}$. In fact, we find that assuming $V_b/\sqrt{\hat gd}\propto\mu_b$ for fully impact-sustained conditions results in a significantly improved data collapse [Fig.~\ref{Slipvelocity}(c)], which makes sense because one can expect that impact entrainment is the more difficult (larger $V_b/\sqrt{\hat gd}$) the larger the granular resistance at the bed surface (larger $\mu_b$). Note that Fig.~\ref{Slipvelocity}(c) corresponds to Fig.~\ref{Slipvelocity}(b) when setting $\mu_b=0.8=\mathrm{const}$.

The transition to a fully impact-sustained transport regime is determined by two independent sufficient conditions. First, the impact number has to exceed a critical value: $\mathrm{Im}\gtrsim20$ (Fig.~\ref{Slipvelocity}, open symbols). This follows from the fact that the transport-layer average $\overline{v_x}$ of the horizontal particle velocity must be larger than the bed velocity as $\langle v_x\rangle(z)$ increases with $z$. For relatively viscous conditions at grain scale ($\mathrm{Ga} < 5$) and close to the entrainment threshold ($\Theta^e_t$), the scaling of the average particle velocity $\overline{v_x}$ can be obtained from the proportionality of $\overline{v_x}$ with the transport-layer-averaged fluid velocity $\overline{u_x}$ (Fig.~\ref{Meanvelocity}) and a dynamic-friction condition [i.e., $\overline{u_x}-\overline{v_x}\propto\mathrm{Ga}\sqrt{(s-1)gd}$ \cite{PaehtzDuran17b}]. Thus, $\overline{v_x} \propto \mathrm{Ga}\sqrt{(s-1)gd} = \mathrm{Im} \sqrt{\hat gd}$ and the condition $\overline{v_x}>V_b \approx \sqrt{\hat gd}$ implies $\mathrm{Im} > \mathrm{Im}_c$, with $\mathrm{Im}_c \approx 20$.

The second sufficient condition for fully impact-sustained transport is $\Theta \gtrsim 5/\mathrm{Im}$ and follows from the proportionality of the dimensionless bed velocity with the dimensionless near-bed fluid velocity in the fully fluid-sustained regime ($V_b/\sqrt{\hat gd} \propto \Theta\mathrm{Im}$) and the fact that $V_b/\sqrt{\hat gd}$ cannot increase indefinitely [Figs.~\ref{Slipvelocity}(b) and \ref{Slipvelocity}(c)]. Therefore, for $\mathrm{Im}<20$, which exclusively characterizes viscous bedload transport conditions, increasing the Shields parameter leads to a transition from fully fluid-sustained transport when $\Theta\lesssim1/\mathrm{Im}$ to fully impact-sustained transport when $\Theta\gtrsim5/\mathrm{Im}$ as the increasing bed velocity reaches the maximum value needed to replace every particle trapped at the bed by exactly one particle entrained through impacts [insets of Figs.~\ref{Slipvelocity}(b) and \ref{Slipvelocity}(c)]. Note that, near the threshold, the condition $\Theta\gtrsim 5/\mathrm{Im}$ always implies $\mathrm{Im}\gtrsim20$ (i.e., $\Theta^e_t < 0.2$ \cite{Ouriemietal07}), but not vice versa [e.g., $s=10^7$, $\mathrm{Ga}=0.1$ in Figs.~\ref{Slipvelocity}(b) and \ref{Slipvelocity}(c)], and the condition $\mathrm{Im}\lesssim5$ always implies $\Theta\lesssim1/\mathrm{Im}$.
\begin{figure}[tb]
 \begin{center}
  \includegraphics[width=1.0\columnwidth]{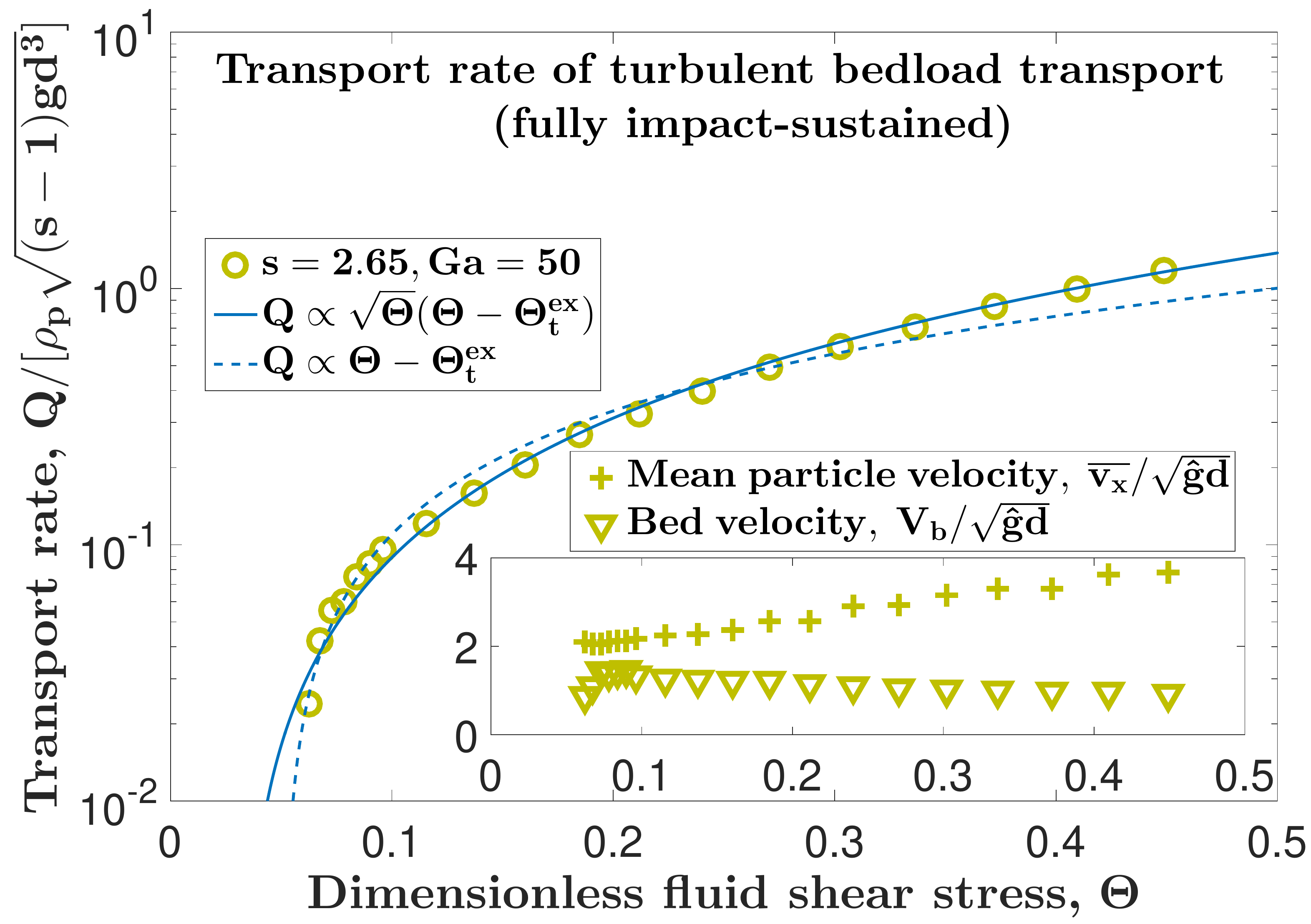}
 \end{center}
 \caption{Dimensionless sediment transport rate $Q/[\rho_p\sqrt{(s-1)gd^3}]$ vs dimensionless fluid shear stress $\Theta$ for fully impact-sustained bedload transport ($s=2.65$ and $\mathrm{Ga}=50$). Inset: Dimensionless transport-layer-averaged horizontal particle velocity $\overline{v_x}/\sqrt{\hat gd}$ and bed velocity $V_b/\sqrt{\hat gd}$ vs $\Theta$.}
 \label{Bedloadtransport}
\end{figure}
\begin{figure*}[htb]
 \begin{center}
  \includegraphics[width=2.0\columnwidth]{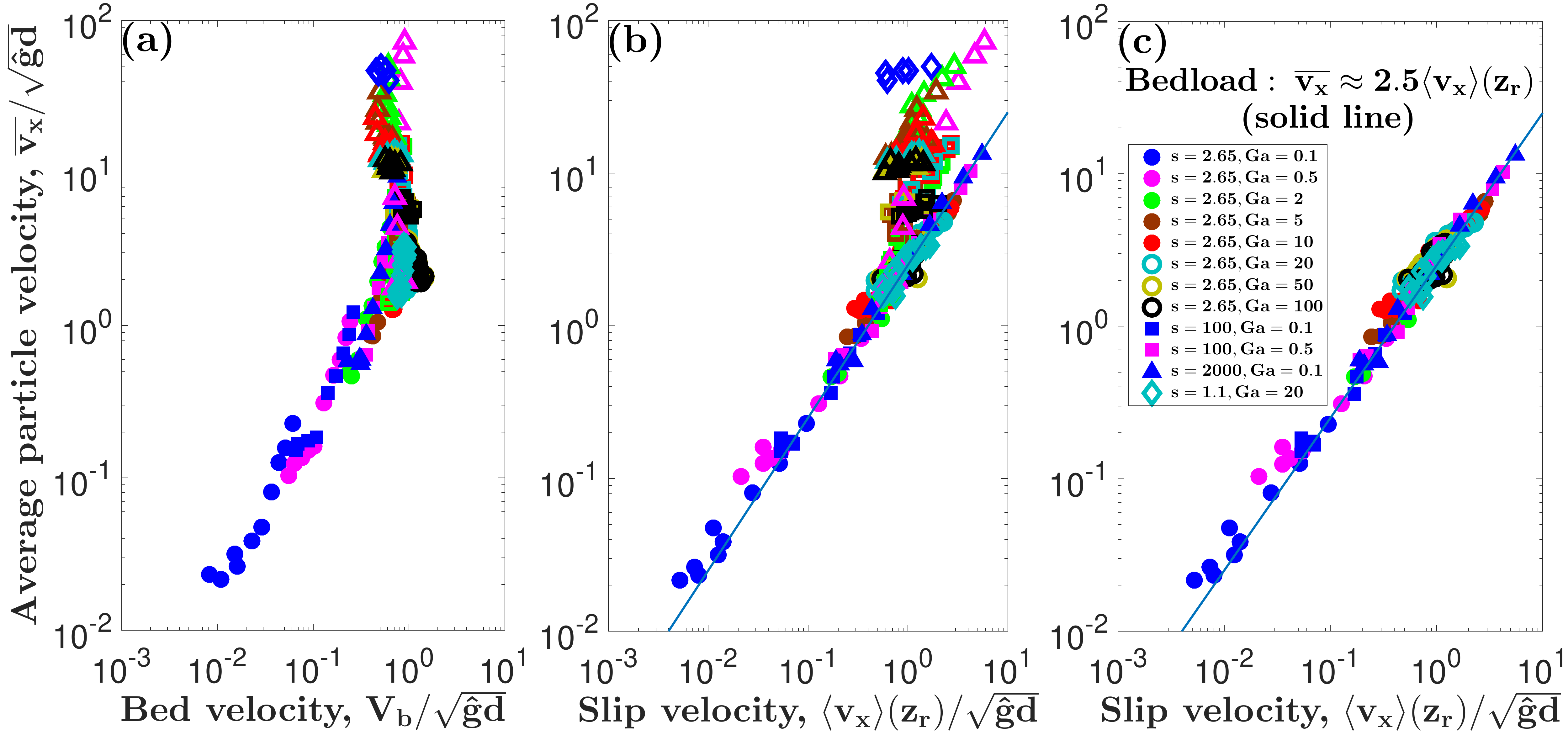}
 \end{center}
 \caption{Correlation between average horizontal particle velocity $\overline{v_x}$ and (a) bed velocity $V_b$ and (b) slip velocity $\langle v_x\rangle(z_r)$. (c) Same as (b), but only for bedload transport conditions. For symbol legend, see Fig.~\ref{Velprofiles}.}
 \label{Vx_vs_Vb}
\end{figure*}

\section{Link between bed velocity and average transport characteristics} \label{Implications}
It is commonly argued that the average horizontal particle velocity $\overline{v_x}$ in the fully impact-sustained regime is constant \cite{UngarHaff87,JenkinsValance14,Berzietal16,Berzietal17} or nearly constant \cite{Andreotti04,Kok10a,Kok10b,Paehtzetal12,Laemmeletal12}, and that the sediment transport rate therefore approximately scales as $Q\propto\Theta-\Theta^r_t$, because the slip velocity is constant. In light of our finding of a generally non-constant slip velocity, but constant bed velocity $V_b$, in the fully impact-sustained regime (Fig.~\ref{Slipvelocity}), one should actually rephrase this argument and say that $\overline{v_x}$ is constant because the bed velocity $V_b$ is constant in this regime. However, even when rephrased, we find that this argument is not valid for fully impact-sustained bedload transport (Fig.~\ref{Bedloadtransport}). Although $Q$ is, indeed, linear in $\Theta$ when $\Theta\lesssim 2\Theta^r_t$, it transforms into a $\Theta^{1.5}$-dependency when $\Theta\gtrsim 2\Theta^r_t$. As a consequence, the scaling $Q\propto\sqrt{\Theta}(\Theta-\Theta^r_t)$, which is consistent with measurements of the bedload transport rate \cite{Smart84,Lajeunesseetal10,CapartFracarollo11}, provides a much better overall fit to the simulations.

The transition from a linear to a non-linear transport law for turbulent bedload transport is consistent with a transition from a constant average particle velocity $\overline{v_x}$ to one that increases with $\Theta$, which occurs even though $V_b$ remains nearly constant (inset of Fig.~\ref{Bedloadtransport}). A similar transition in $\overline{v_x}(\Theta)$ can be found for some other fully impact-sustained conditions [Fig.~\ref{Softbed}(b)] and a similar lack of correlation of $\overline{v_x}$ with $V_b$ for all fully impact-sustained conditions [Fig.~\ref{Vx_vs_Vb}(a)]. Also the classical slip velocity $\langle v_x\rangle(z_r)$ usually does not correlate with $\overline{v_x}$ [Fig.~\ref{Vx_vs_Vb}(b)] with the exception of bedload transport conditions [Fig.~\ref{Vx_vs_Vb}(c)].

\begin{figure}[tb]
 \begin{center}
  \includegraphics[width=1.0\columnwidth]{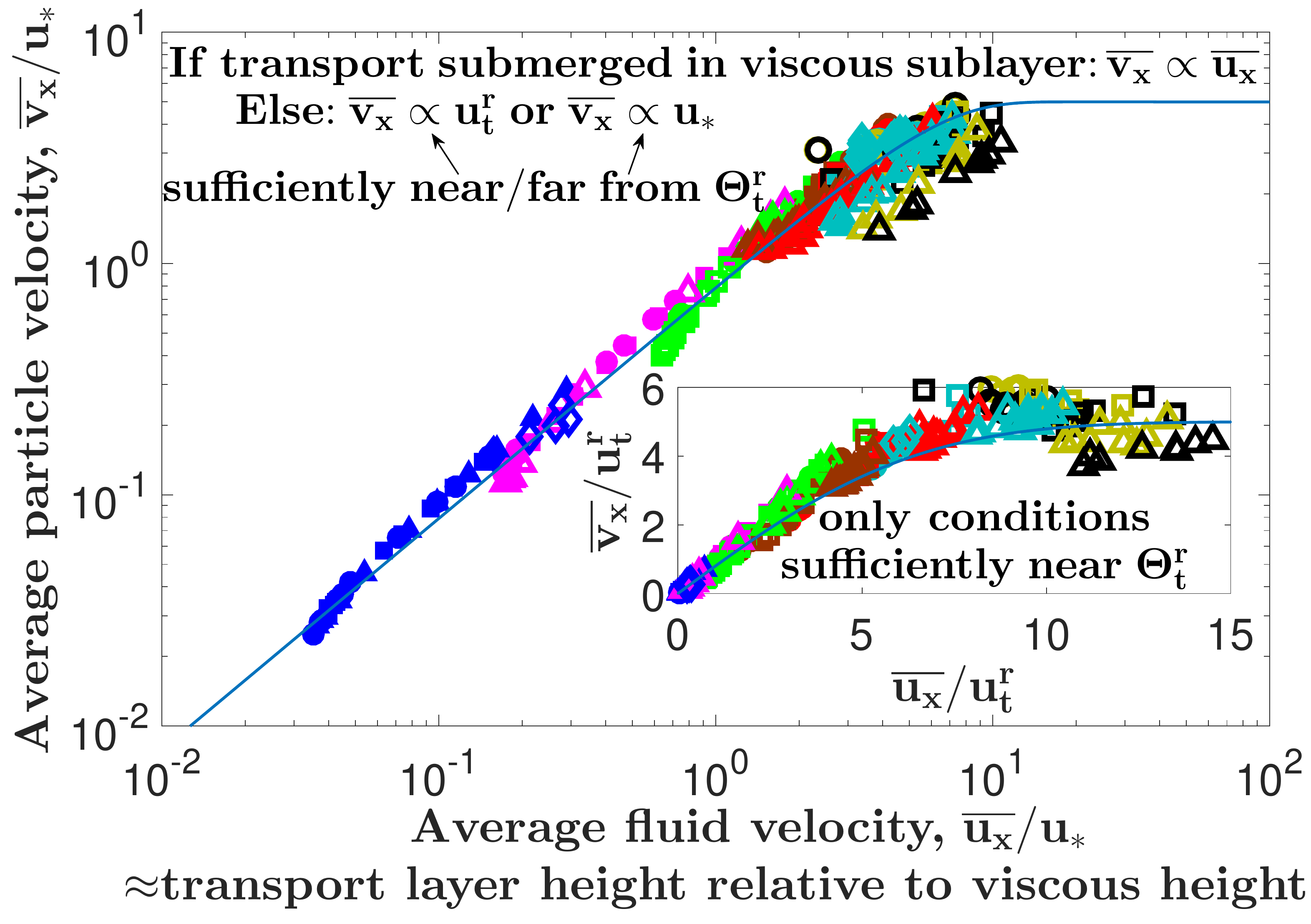}
 \end{center}
 \caption{Average particle velocity $\overline{v_x}$ vs average fluid velocity $\overline{u_x}$, both rescaled by the fluid shear velocity $u_\ast$, for various $s$, $\mathrm{Ga}$, and $\Theta$. Inset: The same in linear-linear scale, but both velocities rescaled by the fluid shear velocity at the threshold ($u^r_t$). Among the fully impact-sustained simulation cases, only those corresponding to the linear transport rate regime are shown in the inset (see text). The solid lines correspond to $\overline{v_x}/u^{\mathrm{typ}}_\ast=5.0\sqrt{1-\exp\left[-0.025\left(\overline{u_x}/u^{\mathrm{typ}}_\ast\right)^2\right]}$, where $u^{\mathrm{typ}}_\ast=u_\ast$ (main figure) or $u^{\mathrm{typ}}_\ast=u^r_t$ (inset). For symbol legend, see Fig.~\ref{Velprofiles}.}

 \label{Meanvelocity}
\end{figure}

\begin{figure}[htb]
 \begin{center}
  \includegraphics[width=1.0\columnwidth]{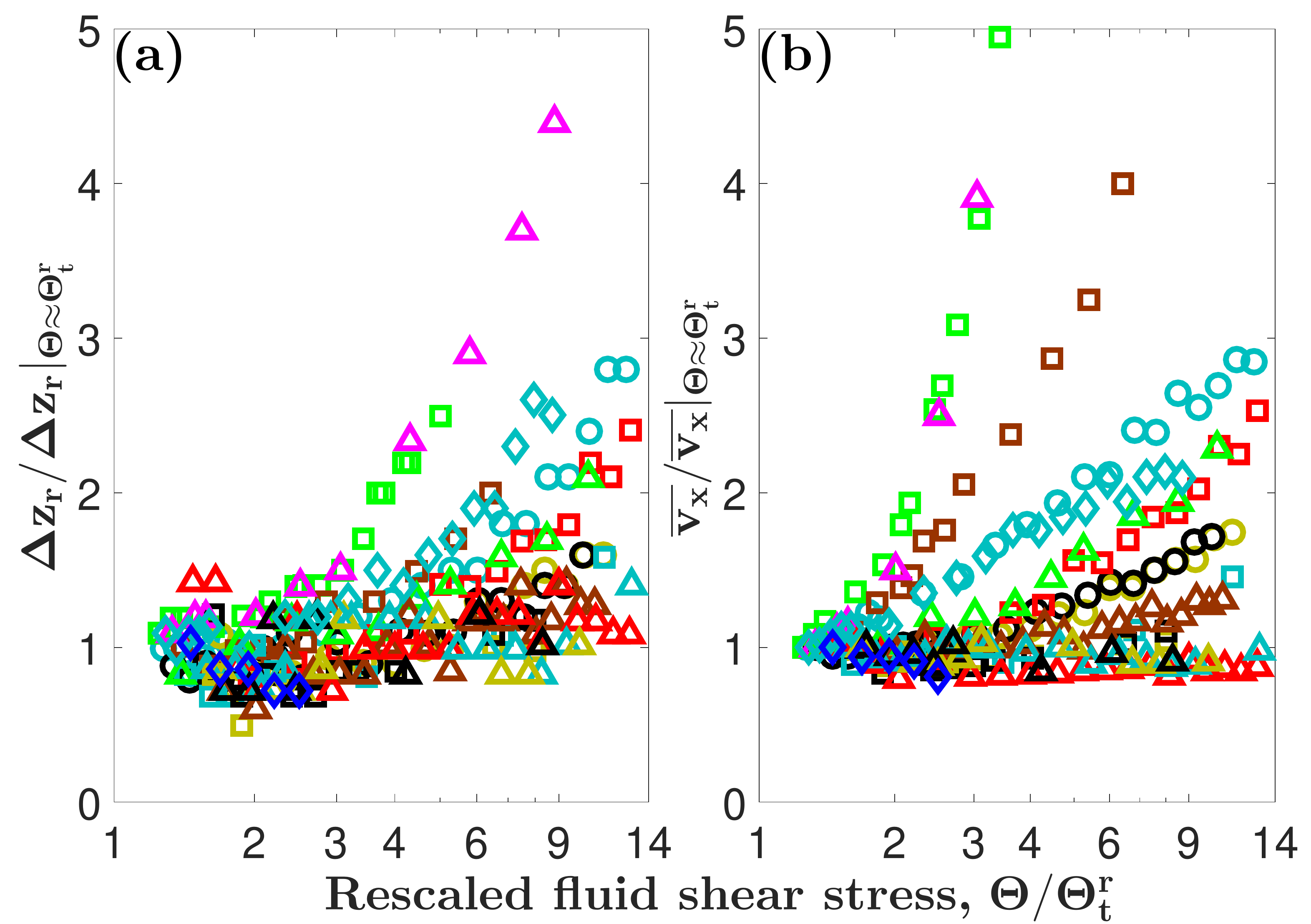}
 \end{center}
 \caption{(a) Rescaled effective location of energetic rebounds relative to the quasistatic bed $\Delta z_r/\Delta z_r|_{\Theta\approx\Theta^r_t}$ vs rescaled dimensionless fluid shear stress $\Theta/\Theta^r_t$ for fully impact-sustained conditions. (b) Rescaled average particle velocity $\overline{v_x}/\overline{v_x}|_{\Theta\approx\Theta^r_t}$ vs $\Theta/\Theta^r_t$ for fully impact-sustained conditions. For symbol legend, see Fig.~\ref{Velprofiles}.}
 \label{Softbed}
\end{figure}

Rather than with the bed velocity, the horizontal particle velocity scales with the horizontal fluid velocity (Fig.~\ref{Meanvelocity}). In detail, we find that the scaling of $\overline{v_x}$ depends on the relation between $\overline{v_x}$ and the size of the transport layer ($\overline{z}-z_r$). When the transport layer is within the viscous sublayer of the turbulent boundary layer [$(\overline{z}-z_r)/z_\nu \lesssim 5$, with the viscous length $z_\nu \equiv d/[\sqrt{\Theta}\mathrm{Ga}]$], $\overline{v_x}$ scales with the characteristic fluid velocity within the viscous sublayer: $\overline{v_x}\propto\overline{u_x}$, where $\overline{u_x}$ scales as $\overline{u_x}\approx\sqrt{\Theta(s-1) g d}(\overline{z}-z_r)/z_\nu$ when the particle-flow feedback (see below) can be neglected. On the other hand, when the transport layer extends beyond the viscous sublayer, the scale of the particle velocity is dominated by the characteristic fluid velocity in the logarithmic region of the velocity profile. That is, $\overline{v_x}\propto\sqrt{\Theta(s-1)gd}\equiv u_\ast$ when the particle-flow feedback can be neglected.

However, for saltation transport in Earth's atmosphere ($s=2000$, $\mathrm{Ga} \gtrsim 10$), it is well known that the particle-flow feedback cannot be neglected because there is a strong drag on the flow generated by particle motion \cite{Bagnold36,Bagnold38,Rasmussenetal96}, which is a necessary condition to maintain a constant average impact velocity \cite{Duranetal11,Koketal12}. As a consequence, the fluid shear velocity remains approximately constant with $\Theta$ in an extended region above the bed surface, and the particle velocity scales as $\overline{v_x} \propto \sqrt{\Theta^r_t(s-1)gd}\equiv u^r_t$ (inset of Fig.~\ref{Meanvelocity}), resulting in a linear scaling of $Q$ with $\Theta$, consistent with measurements \cite{Creysselsetal09,Hoetal11,MartinKok17,Mayaudetal17}. We propose that the same negative feedback keeps a constant average particle velocity, at least close enough to the threshold, in all fully impact-sustained regimes (inset of Fig.~\ref{Meanvelocity}), including turbulent bedload transport. However, the simulations suggest that, at sufficiently large fluid shear stresses ($\Theta\gtrsim2\Theta^r_t$ for turbulent bedload transport), a highly collisional layer of transported particles (a liquidlike or ``soft'' bed \cite{Carneiroetal13}) develops as the bed surface becomes completely mobile (``stage-3'' bedload transport \cite{FreyChurch11}, Movie~S4 \cite{Suppl_Mat}). This liquidlike bed hinders particles moving over it from reaching the disturbed-flow region near the quasistatic-bed surface as they tend to rebound from the liquid-bed surface (Movie~S4 \cite{Suppl_Mat}). Hence, these particles can remain extended periods of time in the nearly undisturbed-flow region, leading to an increase of $\overline{v_x}$ with $\Theta$. This point is further supported by Fig.~\ref{Softbed}(a), which shows that, for fully impact-sustained conditions, the beginning increase of $\overline{v_x}$ [Fig.~\ref{Softbed}(b)] approximately coincides with a beginning increase of the effective location $z_r$ of energetic particle rebounds relative to the quasistatic bed location $z_s$:
\begin{eqnarray}
 \Delta z_r=z_r-z_s,
\end{eqnarray}
where $z_s$ is defined through
\begin{eqnarray}
 \mu(z_s)=0.7\mu_b.
\end{eqnarray}
This definition accounts for potential dependencies of $z_s$ on the contact friction coefficient $\mu^c$. For a typical value $\mu_b=0.6$, it corresponds to $\mu(z_s)=0.42$, consistent with the definition of the quasistatic bed surface applied in previous studies of bedload transport \cite{Maurinetal15}.


\section{Discussions and Conclusions} \label{Conclusions}
Our study challenges the paradigm that sediment transport mediated by water \cite{Paphitis01,FreyChurch11,Houssaisetal15} or heavy air \cite{Koketal12,Burretal15,Berzietal16,Berzietal17}, like on Venus and Titan, is sustained through direct fluid entrainment of bed particles. Using direct sediment transport simulations in a Newtonian fluid for a wide range of the particle-fluid-density ratio $s$, Galileo number $\mathrm{Ga}$, and Shields number $\Theta$, we have shown that the effective horizontal near-bed particle velocity (`bed velocity') in natural units ($V_b/\sqrt{\hat gd}$) becomes a universal constant when the `impact number' $\mathrm{Im}=\mathrm{Ga}\sqrt{s+0.5}\gtrsim20$ or $\Theta\gtrsim 5/\mathrm{Im}$ [Figs.~\ref{Slipvelocity}(b) and \ref{Slipvelocity}(c)]. This result indicates that sediment transport is sustained solely through particle-bed impacts when $\mathrm{Im}\gtrsim20$, which includes nearly all relevant sediment transport regimes. Only sufficiently viscous bedload transport at grain scale is partially ($\mathrm{Im}\lesssim20$ and $\Theta\lesssim 5/\mathrm{Im}$) or fully ($\mathrm{Im}\lesssim20$ and $\Theta\lesssim 1/\mathrm{Im}$) sustained through direct fluid entrainment. However, visualizations of the simulations indicate that the quality of impact entrainment in turbulent bedload is quite different from the one in saltation transport, known as ``splash''. While in saltation transport, the entrained particles are literally ejected from the bed, in bedload transport they are rather dragged out of their traps by the impacting particles (Fig.~\ref{Visualizations} and Movies~S1-S3 \cite{Suppl_Mat}). Note that, for fully impact-sustained transport, the impact number scales as $\mathrm{Im}=(s+0.5)d\sqrt{\hat gd}/\nu\propto(s+0.5)dV_b/\nu$ and may therefore be interpreted as a Stokes number associated with particle-bed impacts.

Our study further challenges the very common assumption in saltation transport modeling that the entire particle motion can be represented by particles moving in identical periodic trajectories \cite{Bagnold41,Kawamura51,Owen64,Kind76,LettauLettau78,UngarHaff87,Sorensen91,Sorensen04,Sauermannetal01,ClaudinAndreotti06,DuranHerrmann06,Duranetal11,JenkinsValance14,Berzietal16,Berzietal17}. If this assumption was true, the transport-layer-averaged horizontal particle velocity $\overline{v_x}$ would be bounded between the horizontal velocities at take-off and impact and thus be approximately proportional to the slip velocity $\langle v_x\rangle(z_r)$, which is the average particle velocity at the location of the bed surface ($z_r$). However, we find that $\overline{v_x}$ scales with the fluid velocity within the transport layer (Fig.~\ref{Meanvelocity}) and generally not with $\langle v_x\rangle(z_r)$ nor $V_b$ (Fig.~\ref{Vx_vs_Vb}). We also find that the locally averaged vertical particle velocity $\sqrt{\langle v_z^2\rangle}(z)$ increases exponentially with elevation $z$ near the bed surface [Fig.~\ref{Velprofiles}(b)], whereas an identical-trajectory model necessarily predicts a decrease. These discrepancies are evidence for a separation of particle velocity scales, which was already pointed out for saltation transport in a previous study (Fig.~21 of Ref.~\cite{Duranetal11}). Near the bed surface, the average particle velocity is dominated by a comparably slow species of particle (`reptons' \cite{Andreotti04}, or ``leapers'' and ``creepers'' \cite{Carneiroetal13}), whereas at larger elevations that cannot be reached by the slow species, a comparably fast species dominates (``saltons'' \cite{Andreotti04,Carneiroetal13}). This being said, identical-trajectory representations of sediment transport do have their uses. For instance, they seem to give valuable insights into the physics of sediment transport cessation \cite{Berzietal16,Berzietal17,PaehtzDuran17b}.

Finally, for fully impact-sustained transport, our study predicts a relatively strong negative feedback of the particle motion on the flow when the dimensionless fluid shear stress is sufficiently close to $\Theta^r_t$. As a consequence, $\overline{v_x}$, which is controlled by the flow, remains approximately constant with $\Theta$, leading to a linear scaling of the sediment transport rate ($Q\propto\Theta-\Theta^r_t$). However, for turbulent bedload transport, this linear scaling becomes non-linear slightly above the threshold ($\Theta\approx2\Theta^r_t$, see Fig.~\ref{Bedloadtransport}) due to a sudden drop in the relative feedback strength, which is associated with the formation of a liquid-like bed of particles on top of the quasistatic bed surface (Fig.~\ref{Softbed}).

Our numerical finding that steady turbulent bedload transport is fully sustained through entrainment by particle-bed impacts may be criticized because the simulations neglect a number of items that are deemed to have a significant influence on bedload transport: they neglect the hindrance effect (i.e., an increase of the average fluid drag force at large particle concentrations), are quasi-two-dimensional, and only account for the mean turbulent flow, but not for turbulent fluctuations around the mean, which are known to be crucial for the initiation of bedload transport \cite{Diplasetal08,Valyrakisetal10,Valyrakisetal13}. However, we believe that our finding is robust because our simulations quantitatively reproduce measurements of bedload transport cessation thresholds \cite{PaehtzDuran17b}, which would not be expected if these neglected items played a crucial role for sustaining steady bedload transport. Our reasoning is supported by Ref.~\cite{Maurinetal15}, who compared three-dimensional bedload transport simulations with and without turbulent fluctuations. Their Fig.~6 indicates that, although the initiation threshold is strongly affected by turbulent fluctuations, the cessation threshold is nearly unaffected because the extrapolation of the simulated transport rates to vanishing transport remains nearly the same.

\appendix
\section{Movie captions} \label{Movies}
\subsection{Movie~S1}
Time evolution of the simulated particle-fluid system for $s=2000$, $\mathrm{Ga}=20$, and $\Theta\simeq2.1\Theta^r_t$, considering weakly-damped binary collisions ($e=0.9$). The flow velocity is shown as a background color with warm colors corresponding to high velocities and cold colors to small velocities. The horizontal and vertical axes are measured in mean particle diameters. Only $1/4$ of the simulated horizontal domain is shown, which is why there are occasions at which no moving particle can be observed. This is an example for saltation transport, which is predominantly sustained through particle-bed impact entrainment. One can see that impacting particles tend to eject surface particles.

\subsection{Movie~S2}
Time evolution of the simulated particle-fluid system for $s=2.65$, $\mathrm{Ga}=20$, and $\Theta\simeq1.7\Theta^r_t$, considering binary collisions that are nearly fully damped by the lubrication force ($e=0.01$). The flow velocity is shown as a background color with warm colors corresponding to high velocities and cold colors to small velocities. The horizontal and vertical axes are measured in mean particle diameters. Only $1/4$ of the simulated horizontal domain is shown, which is why there are occasions at which no moving particle can be observed. This is an example for turbulent bedload transport, which is predominantly sustained through particle-bed impact entrainment. One can see that impacting particles tend to drag surface particles out of they traps.

\subsection{Movie~S3}
Time evolution of the simulated particle-fluid system for $s=2000$, $\mathrm{Ga}=20$, and $\Theta\simeq 17.1\Theta^r_t$, considering weakly-damped binary collisions ($e=0.9$). The flow velocity is shown as a background color with warm colors corresponding to high velocities and cold colors to small velocities. The horizontal and vertical axes are measured in mean particle diameters. Only $1/4$ of the simulated horizontal domain is shown, which is why there are occasions at which no moving particle can be observed. This is an example for saltation transport, which is predominantly sustained through particle-bed impact entrainment. One can see that impacting particles tend to eject surface particles.

\subsection{Movie~S4}
Time evolution of the simulated particle-fluid system for $s=2.65$, $\mathrm{Ga}=20$, and $\Theta\simeq 13.8\Theta^r_t$, considering binary collisions that are nearly fully damped by the lubrication force ($e=0.01$). The flow velocity is shown as a background color with warm colors corresponding to high velocities and cold colors to small velocities. The horizontal and vertical axes are measured in mean particle diameters. Only $1/4$ of the simulated horizontal domain is shown. This is an example for turbulent bedload transport, which is predominantly sustained through particle-bed impact entrainment. However, it is impossible to determine single entrainment events because several layers of the bed are in continuous motion. These layers constitute the liquid-like bed, and it can be seen that energetic particles tend to rebound from its top.

\begin{acknowledgments}
We acknowledge support from the National Natural Science Foundation of China (Grant No. 11550110179).
\end{acknowledgments}

%

\end{document}